\documentclass[12pt]{article}
\usepackage{amsfonts,amssymb,amsmath,amsthm}
\usepackage{epsfig}
\usepackage{slashed}
\usepackage{graphicx}
\usepackage{mathrsfs}
\usepackage{bbm}
\usepackage{color}
\def\and{{\rm and}}

\def\a{\alpha}
\def\b{\beta}
\def\e{\eta}
\def\ep{\epsilon}
\def\d{\delta}
\def\g{\gamma}

\def\l{\lambda}
\def\s{\sigma}

\def\om{\omega}
\def\p{\partial}
\def\th{\theta}

\def\L{\Lambda}
\def\O{\Omega}

\begin{document}
\renewcommand{\thefootnote}{\fnsymbol{footnote}}
\begin{titlepage}

\vspace{10mm}
\begin{center}
{\Large\bf $SL(2,C)$ gravity on noncommutative space with Poisson structure}
\vspace{16mm}

{\large Yan-Gang Miao\footnote{\em E-mail: miaoyg@nankai.edu.cn}
and Shao-Jun Zhang\footnote{\em E-mail: sjzhang@mail.nankai.edu.cn}

\vspace{6mm}
{\normalsize \em Department of Physics, Nankai University, Tianjin 300071, \\
People's Republic of China}}

\end{center}

\vspace{10mm}
\centerline{{\bf{Abstract}}}
\vspace{6mm}

The Einstein's gravity theory can be formulated as an $SL(2,C)$ gauge theory in terms of spinor notations. In this paper, we consider a
noncommutative space with the Poisson structure and construct an $SL(2,C)$ formulation of gravity on such a space. Using the covariant coordinate
technique, we build a gauge invariant action in which, according to the Seiberg-Witten map, the physical degrees of freedom are expressed in terms of
their commutative counterparts up to the first order in noncommutative parameters.

\vskip 20pt
PACS Number(s): 04.90.+e; 02.40.Gh; 11.10.Nx


\end{titlepage}

\newpage

\pagenumbering{arabic}

\section{Introduction}
In recent years, the idea of noncommutative spacetimes has attracted much attention although it was proposed by Synder~\cite{Snyder}
as early as in 1947 in order to remove the divergence in quantum field theories. It has been argued that at a very small scale, say the Planck length,
coordinates of spacetimes cannot be measured at any accuracy~\cite{DFR}, i.e. the measurement should satisfy a set of uncertainty relations that
can be well realized within the context of noncommutative sapcetimes. Starting from string theory, Seiberg and Witten~\cite{SW}
suggested that the D-brane dynamics with a $B$-field background can be described by some noncommutative field theory.
Recently
the idea of noncommutative spacetimes has penetrated into various fields in physics. The research on the construction of quantum field theories
on noncommutative spacetimes is fruitful (for reviews, see ref.~\cite{NCQFT}) and it is remarkable to note that noncommutative quantum field theories
can be applied to the study related to strong background fields, such as the quantum Hall effect (for a review, see ref.~\cite{NCBP}).

It is interesting to study gravity on noncommutative spacetimes. Actually, the noncommutative formulation of gravity has been
considered~\cite{NCgravrev} to be
a necessity for quantization of gravity. The main obstacle for this formulation is on dealing with
the general coordinate invariance.
Recently there have been some approaches proposed for solving this problem. In ref.~\cite{Chamseddine:2000}
a deformation of Einstein's gravity is constructed
based on gauging the noncommutative $SO(4,1)$ de Sitter group and then contracting it to $ISO(3,1)$
in terms of the Seiberg-Witten map~\cite{SW}. Another effort~\cite{Marculescu:2008} also from the point of view of the Seiberg-Witten map
is made to build the $SO(3,1)$ noncommutative formulation of gravity.
In refs.~\cite{Chamseddine:2002,Cardellaetal:2002} noncommutative gravity models are established by the reduction of the constrained  $U(2,2)$
to $SO(3,1)$.
Moreover, the theory of gravity can also be expressed in a $GL(2,C)$ formulation with complex vierbeins~\cite{Chamseddine:2003}.
Within the framework of the gauge theory of gravity, the authors of ref.~\cite{Calmet:2005} have given
a noncommutative formulation of
gravity based on a class of restricted diffeomorphism symmetries that preserves the noncommutative algebra.
On the other hand, Wess and collaborators~\cite{TDG} have proposed a gravitational theory considering a twisted diffeomorphism algebra
from a purely geometrical point of view.

The formulations of noncommutative gravity mentioned above are established in the canonical noncommutative spacetime in which
coordinates $\hat{x}^\mu$ satisfy the following commutation relations,
\begin{equation}
[\hat{x}^\mu, \hat{x}^\nu] = i \th^{\mu\nu},
\end{equation}
where $\th^{\mu\nu}$ is an antisymmetric constant tensor and its elements are called noncommutative parameters.
The noncommutativity can be realized in the ordinary spacetime
with the replacement of the ordinary product by the star-product between functions. It is reasonable to consider a more general noncommutative spacetime
where noncommutative parameters are coordinate-dependent. For the most general noncommutative parameters $\th^{\mu\nu}(\hat{x})$,
the star-product between functions may
become non-associative, which gives rise to quite complicated problems. Fortunately, there exists a class of noncommutative manifolds on which
the star-product between functions is associative, like the Poisson manifold. In the present paper, we focus our attention on such a manifold.
There are already some works that deal with gravity on coordinate-dependent noncommutative spacetimes. In ref.~\cite{Banerjee:2007}
the theory of gravity is constructed based on the work of ref.~\cite{Calmet:2005} on a noncommutative spacetime with the Lie algebraic structure,
which is in fact a special case of the Poisson manifold. In terms of the twisted differential geometry,
another theory of gravity is proposed~\cite{Aschieri:2009} which is invariant under diffeomorphism
as well as under a $GL(2,C)$ $\star$-gauge transformation on a general noncommutative spacetime. We prefer to carry out our analysis
in the framework of the $SL(2,C)$
formulation~\cite{Ishametal:1972,Chamseddine:2003}. The advantage of this formulation lies in a tetrad formalism where the tetrad transforms
covariantly under gauge transformations, so that we can construct a gauge invariant action in a natural way. As a gauge theory, it enables one to use
the machinery of noncommutative gauge theories elaborately developed in the
literature~\cite{Wessetal:20001,Wessetal:20002,Wessetal:2001,Calmet:2003}. In terms of the covariant coordinate technique~\cite{Wessetal:20001}
a rank two tensor $\hat{R}^{\mu\nu}$ is constructed at first. Because of the coordinate-dependence of $\th^{\mu\nu}({\hat x})$,
it is not straightforward to write the covariantly transformed curvature tensor $\hat{R}_{\mu\nu}$ through the relation
$\hat{R}^{\mu\nu} = \th^{\mu\l} \th^{\nu\s} \hat{R}_{\l\s}$ as was done in the canonical noncommutative spacetime. To this end, a modified function
$\hat{\th}_{\mu\nu}$ (see eq.~(\ref{NCRrelation}) and eq.~(\ref{deltatheta})) is introduced in order for the gauge field strength
to transform covariantly.
As a result, the action can be constructed in terms of the curvature tensor $\hat{R}_{\mu\nu}$ and the vierbein
$\hat{e}_\mu$. Furthermore, the Seiberg-Witten map~\cite{SW} in our case is derived for noncommutative physical quantities up to
the first order in the coordinate-dependent noncommutative parameters,
we can therefore express the noncommutative theory in terms of ordinary physical quantities completely.

The paper is organized as follows. In the next section we give a brief introduction to the $SL(2,C)$ formulation of gravity
on the ordinary spacetime. In the first subsection
of section 3, the formulation is extended to a noncommutative space with the Poisson structure and a gauge invariant action is thus constructed.
In the second subsection, the Seiberg-Witten map of the noncommutative formulation is derived up to the first order in the
coordinate-dependent noncommutative parameters.
The last section is devoted to conclusion. As to notations, we use the Latin letters, $a,b,\cdots = 0,1,2,3$,
to denote Lorentz indices and the Greek letters, $\mu,\nu,\cdots = 0,1,2,3$, spacetime indices.

\section{A brief introduction to $SL(2,C)$ formualtion of gravity}
Before discussing its noncommutative formulation, let us recall briefly the $SL(2,C)$ formulation of
gravity~\cite{Ishametal:1972,Chamseddine:2003,Stern:2009}
on the ordinary spacetime.

There are some physical quantities we need to construct the action of gravity. At first the $SL(2,C)$ gauge field $\om$ is introduced as
\begin{equation}
\om = \frac{1}{2} \om_\mu^{~ab} \s_{ab} dx^\mu = \om_\mu dx^\mu,
\end{equation}
where $\s_{ab} = - \frac{i}{4} [\g_a, \g_b]$ are $SL(2,C)$ generators and $\g_a$ are Dirac gamma matrices satisfying the anticommutation relations
$\{\g_a, \g_b\} = 2 \e_{ab}$.
Then the curvature tensor is given in terms of $\om$ from its definition,
\begin{equation}
R \equiv \frac{1}{4} R_{\mu\nu}^{~~ab} \s_{ab} dx^\mu \wedge dx^\nu = d \om - i \om \wedge \om. \label{Curvature}
\end{equation}
In addition, the vierbein $e$ is introduced as follows,
\begin{equation}
e= e^a_\mu \g_a d x^\mu = e_\mu d x^\mu.
\end{equation}

Under the $SL(2,C)$ transformation, $\om$ and $e$ transform as
\begin{eqnarray}
&&e \rightarrow \O e \O^{-1} \label{e_transformation},\\
&&\om \rightarrow \O \om \O^{-1} + i \O d \O^{-1}\label{om_transformation},
\end{eqnarray}
where the transformation parameter $\O = \exp(i \frac{1}{2} \Lambda^{ab} \s_{ab}) \equiv \exp(i \Lambda)$.
In infinitesimal forms, eq.~(\ref{e_transformation}) and eq.~(\ref{om_transformation}) can be written as
\begin{eqnarray}
\d_{\L} e &=& i [\L, e] \label{Vierbtran},\\
\d_{\L} \om &=& d \L + i [\L, \om] \label{Spintran}.
\end{eqnarray}
Using eq.(\ref{Curvature}) and eq.~(\ref{om_transformation}), one can show that the curvature tensor $R$ transforms covariantly
under the $SL(2,C)$ gauge transformation,
\begin{equation}
R \rightarrow \O R \O^{-1},
\end{equation}
where its infinitesimal form is
\begin{equation}
\d_{\L} R = i [\L, R]. \label{Rtrans}
\end{equation}

Now it is straightforward to write an $SL(2,C)$ invariant action
\begin{eqnarray}
S &=& \int_M \mathrm{Tr} \bigg((c_0 + c_1 \g_5) e \wedge e \wedge R + c_2 {\gamma}_5 e \wedge e \wedge e \wedge e\bigg), \nonumber\\
&=& \int_M  d^4 x ~ \ep^{\mu\nu\rho\s} \mathrm{Tr} \bigg((c_0 + c_1 \g_5) e_\mu  e_\nu  R_{\rho\s}
+ c_2 {\gamma}_5 e_\mu  e_\nu  e_\rho e_\s\bigg),\label{action}
\end{eqnarray}
where $\g_5 = i \g_0 \g_1 \g_2 \g_3$, $c_0$, $c_1$ and $c_2$ are arbitrary constants. Integrating out $\om$ and endowing appropriate values
for $c_0$, $c_1$ and $c_2$ in eq.~(\ref{action}), one can obtain the Einstein-Hilbert action plus a cosmological constant.

In the next section, we generalize this formulation of gravity to a noncommutative space with the Poisson structure. For the sake of convenience,
we do not write actions in forms but in components.

\section{$SL(2,C)$ gravity on noncommutative space}
Consider a noncommutative spacetime whose coordinates satisfy the following commutation relations
\begin{equation}
[\hat{x}^\mu, \hat{x}^\nu] = i \th^{\mu\nu} (\hat{x}) \label{noncommutativity},
\end{equation}
where the coordinate-dependent $\th^{\mu\nu} (\hat{x})$ is a Poisson bivector\footnote{In the sense of the star product (see eq.~(\ref{star}))
the algebraic relations of the noncommutative spacetime can be written as $[{x}^\mu, x^\nu]_\star =i \th^{\mu\nu} (x)$. Thus we can also utilize
$\th^{\mu\nu} (x)$ to denote a Poisson bivector.}
and it can be used to define a Poisson bracket on the manifold,
\begin{equation}
\{f(x), g(x)\}_{\rm Poisson} \equiv \th^{\mu\nu} (x) \p_\mu f(x) \p_\nu g(x),
\end{equation}
where $f(x)$ and $g(x)$ are arbitrary functions on the manifold. The Jacobi indentity of the Poisson bracket imposes the following
conditions on the bivector $\th^{\mu\nu} (x)$,
\begin{equation}
\th^{\mu\rho}(x) \p_\rho \th^{\nu\s}(x) + \th^{\nu\rho}(x) \p_\rho \th^{\s\mu}(x) + \th^{\s\rho}(x) \p_\rho \th^{\mu\nu}(x) = 0 \label{Jacobi}.
\end{equation}
Suppose that the bivector $\th^{\mu\nu} (x)$ is nondegenerate, therefore we can define its inverse $\th_{\mu\nu} (x)$ as:
$\th^{\mu\nu} \th_{\nu\rho} = \d^\mu_\rho$. With the Jacobi identity eq.~(\ref{Jacobi}), we can show that the two-form
$\Theta = \frac{1}{2} \th_{\mu\nu} dx^\mu \wedge dx^\nu$ is closed ($d \Theta=0$) and then the manifold is sympletic.
In this paper, we shall consider only the case in which the manifold is symplectic.

According to Kontsevich's deformation~\cite{Kontsevich:1997}, there exists an associative star product to a given Poisson bivector
$\th^{\mu\nu} (x)$ and it can be written as the following symmetric form up to the first order in $\th^{\mu\nu}(x)$,
\begin{equation}
f(x) \star g(x) = f(x) g(x) + \frac{i}{2} \th^{\mu\nu}(x) \p_\mu f(x) \p_\nu g(x) + \mathcal{O} (\th^2).\label{star}
\end{equation}
Note that it is not unique for higher order terms. In order to avoid this ambiguity we shall restrict our discussion
only to the first order in $\th^{\mu\nu}(x)$.

In the following subsections we construct the $SL(2,C)$ gravity on the spacetime with the structure eq.~(\ref{noncommutativity})
and derive the Seiberg-Witten map of the noncommutative gravity up to the first order in $\th^{\mu\nu}(x)$.

\subsection{Construction of noncommutative gravity}
Let us follow the covariant coordinate approach proposed in ref.~\cite{Wessetal:20001}. The covariant coordinates $\hat{X}^\mu = x^\mu + \hat{B}^\mu$
are defined by the gauge transformation property\footnote{Coordinates are suppressed from subsection 3.1
to the end of this paper for the sake of simplicity.}
\begin{equation}
\d_{\hat{\L}} (\hat{X}^\mu \star \hat{\Psi}) = i \hat{\L} \star (\hat{X}^\mu \star \hat{\Psi}),
\end{equation}
where $\hat{\L}$ is the transformation parameter and $\hat{\Psi}$ is a matter field with the gauge transformation
\begin{equation}
\d_{\hat{\L}} \hat{\Psi} = i \hat{\L} \star \hat{\Psi}.
\end{equation}
This requires that the field $\hat{B}^\mu$ should transform as
\begin{eqnarray}
\d_{\hat{\L}} \hat{B}^\mu &=& i [\hat{\L}, x^\mu]_\star + i [\hat{\L}, \hat{B}^\mu]_\star \nonumber\\
&=& \th^{\mu\nu} \p_\nu \hat{\L} + i [\hat{\L}, \hat{B}^\mu]_\star,
\end{eqnarray}
and the covariant coordinates as
\begin{equation}
\d_{\hat{\L}} \hat{X}^\mu = i [\hat{\L}, \hat{X^\mu}]_\star \label{CovCoorTran}.
\end{equation}
The noncommutative spin-connection $\hat{\om}_\mu$ is given by \cite{Wessetal:20001,Banerjee:2007}
\begin{equation}
\hat{\om}_\mu = \th_{\mu\nu} \hat{B}^\nu \label{Borelation},
\end{equation}
where $\th_{\mu\nu}$ is the inverse of $\th^{\mu\nu}$: $\th_{\mu\nu} \th^{\nu\rho} = \d^\rho_\mu$. Due to the coordinate-dependence of the
noncommutative structure  depicted by $\th^{\mu\nu}$, it is not possible to find the transformation of $\hat{\om}_\mu$ in a closed form
but one can obtain it correct up to any order required in the noncommutative parameter. To the first order in $\th$, we have
\begin{equation}
\d_{\hat{\L}} \hat{\om}_\mu = \p_\mu \hat{\L} + i [\hat{\L}, \hat{\om}_\mu] - \frac{1}{2} \th^{\l\s} \{\p_\l \hat{\L}, \p_\s \hat{\om}_\mu\}
- \frac{1}{2} \th_{\mu\a} \th^{\l\s} \p_\s \th^{\a\b} \{\p_\l \hat{\L}, \hat{\om}_\b\} \label{NCOmtran},
\end{equation}
where $\{\cdot, \cdot\}$ stands for an anticommutator.

Using the covariant coordinates, we can define a rank-two tensor
\begin{equation}
\hat{R}^{\mu\nu} \equiv - i ([\hat{X}^\mu, \hat{X}^\nu]_\star - i \th^{\mu\nu} (\hat{X})) \label{SecTensor},
\end{equation}
and using eq.~(\ref{CovCoorTran}) we find it transforms as follows,
\begin{equation}
\d_{\hat{\L}} \hat{R}^{\mu\nu} = i [\hat{\L}, \hat{R}^{\mu\nu}]_\star.
\end{equation}
Now it is time to look for the relation between the rank-two tensor $\hat{R}^{\mu\nu}$ and the gauge field strength $\hat{R}_{\mu\nu}$.
In the case of canonical noncommutative spaces where $\th$ is constant, the relation is trivial:
$\hat{R}^{\mu\nu} = \th^{\mu\rho} \th^{\nu\s} \hat{R}_{\rho\s}$. But in our case $\th$ is coordinate-dependent, we should modify the relation
and make sure the gauge field strength $\hat{R}_{\mu\nu}$ transform covariantly,
\begin{equation}
\d_{\hat{\L}} \hat{R}_{\mu\nu} = i [\hat{\L}, \hat{R}_{\mu\nu}]_\star \label{NCRtran}.
\end{equation}
Suppose that there exists a function $\hat{\th}_{\mu\nu} (\hat{X})$ which ensures that $\hat{R}_{\mu\nu}$  with the definition
\begin{equation}
\hat{R}_{\mu\nu} \equiv \hat{\th}_{\mu\l} \star \hat{\th}_{\nu\s} \star \hat{R}^{\l\s} \label{NCRrelation}
\end{equation}
satisfies the transformation property eq.~(\ref{NCRtran}). This requires that $\hat{\th}_{\mu\nu} (\hat{X})$ should transform as
\begin{equation}
\d_{\hat{\L}} \hat{\th}_{\mu\nu} (\hat{X}) = i [\hat{\L}, \hat{\th}_{\mu\nu}]_\star.\label{deltatheta}
\end{equation}
In the next subsection, we can see the function $\hat{\th}_{\mu\nu} (\hat{X})$ indeed exists and we shall give its expansion expression.

As in the commutative case, we introduce the noncommutative analogue of vierbeins $\hat{e}_\mu$ with the gauge transformation
\begin{equation}
\d_{\hat{\L}} \hat{e}_\mu = i [\hat{\L}, \hat{e}_\mu]_\star \label{NCe_transformation}.
\end{equation}

Now it is straightforward for us to write a gauge invariant action by using eq.~(\ref{NCRtran}) and eq.~(\ref{NCe_transformation}),
\begin{equation}
S = \int d^4 x \ (\mathrm{det}\th^{\mu\nu})^{-\frac{1}{2}} \ep^{\mu\nu\rho\s} \mathrm{Tr} \bigg((c_0
+ c_1 \g_5)\hat{e}_\mu \star \hat{e}_\nu \star \hat{R}_{\rho\s}
+ c_2 \g_5 \hat{e}_\mu \star \hat{e}_\nu \star \hat{e}_\rho \star \hat{e}_\s \bigg) \label{NCaction}.
\end{equation}
Here the volume form on the symplectic manifold, i.e. $(\mathrm{det}\th^{\mu\nu})^{-\frac{1}{2}}d^4 x $, appears naturally with which the
following trace property of the integral
is satisfied to any order in $\th^{\mu\nu}$,\footnote{If there exists a function $\Omega (x)$ satisfying the relation:
$\partial_\mu (\Omega \theta^{\mu\nu}) = 0$,
then we have the trace property of the integral~\cite{Calmet:2003, Felder:2000}:
$\int d^4 x~\Omega(x) (f(x) \star g(x)) = \int d^4 x~\Omega(x) (g(x) \star f(x))$.
For a symplectic manifold, there exists~\cite{Behr:2003,Vassilevich:2010} a natural choice for the function $\Omega (x)$ which satisfies
the above requirement:
$\Omega= ({\rm det} \theta^{\mu\nu})^{- \frac{1}{2}}$.
Moreover, this is also shown in ref.~\cite{Jurco:2001} from a different point of view.}
\begin{equation}
\int d^4 x(\mathrm{det} \th^{\mu\nu})^{-\frac{1}{2}}\mathrm{Tr} (f \star g)
= \int d^4 x (\mathrm{det}\th^{\mu\nu})^{-\frac{1}{2}} \mathrm{Tr} (g \star f) \label{NCTracepro}.
\end{equation}
It is obvious that the action eq.~(\ref{NCaction}) is indeed gauge invariant because we can verify
\begin{equation}
\d_{\hat{\L}} S = \int d^4 x \ (\mathrm{det}\th^{\mu\nu})^{-\frac{1}{2}} \ep^{\mu\nu\rho\s} \mathrm{Tr} \bigg(i [\hat{\L}, (c_0 + c_1 \g_5)
\hat{e}_\mu \star \hat{e}_\nu \star \hat{R}_{\rho\s} + c_2 \g_5 \hat{e}_\mu \star \hat{e}_\nu \star \hat{e}_\rho \star \hat{e}_\s]_\star\bigg) = 0,
\end{equation}
where eq.~(\ref{NCTracepro}) has been used.

In the next subsection, we connect the noncommutative theory with its commutative counterpart using the so-called ``Seiberg-Witten map''.

\subsection{Seiberg-Witten map to the first order}
A general gauge group with an algebra $G$ does not close on noncommutative spacetimes with the exception to unitary groups.
For consistency, the algebra $G$ should be enlarged to its universal enveloping algebra $\mathcal{U}(G)$~\cite{Wessetal:20002}.

Firstly let us take a close look at the infinitesimal gauge transformation of the spin-connection $\hat{\om}$ in eq.~(\ref{NCOmtran}) where
both commutators and anticommutators appear for the case of noncommutative spacetimes. Secondly note that the commutator
$[\hat{\Lambda}, \hat{e}]_\star$ in the infinitesimal gauge transformation of vierbeins in eq.~(\ref{NCe_transformation}) can be written as
\begin{equation}
[\hat{\Lambda}, \hat{e}]_\star = \frac{1}{4} \{\hat{\Lambda}^{ab}, \hat{e}^c\}_\star [\s_{ab}, \g_c] + \frac{1}{4}
[\hat{\Lambda}^{ab}, \hat{e}^c]_\star \{\s_{ab}, \g_c\}.
\end{equation}
Using the identities of the Dirac gamma matrices,
\begin{eqnarray}
[\s_{ab}, \g_c] &=& i (\e_{ac} \g_b - \e_{bc} \g_a), \\
\{\s_{ab}, \g_c\} &=& - \ep_{abc}^{~~~d}\g_5\g_d, \\
\{\s_{ab}, \s_{cd}\} &=& \frac{1}{2} (i \ep_{abcd} \g_5 + \e_{ac} \e_{bd} - \e_{ad} \e_{bc}),
\end{eqnarray}
we can see that $SL(2,C)$ does not close on noncommutative spacetimes and it should be enlarged to a bigger gauge group including the additional
generators $1$ and $\g_5$, i.e.  $SL(2,C)$ is enlarged to $GL(2,C)$, and that the vierbeins should be extended to include
the additional generator $\g_5\g_a$.
Thus the $GL(2,C)$ spin-connection $\hat{\om}_\mu$ and gauge parameter $\hat{\L}$ can be decomposed as
\begin{equation}
\hat{\om}_\mu = \frac{1}{2} \hat{\om}^{(0) ab}_\mu \s_{ab} + \hat{a}^{(1)}_\mu + i \hat{b}^{(1)}_{\mu 5} \g_5, \ \ \ \ \
\hat{\L} = \frac{1}{2} \hat{\L}^{(0) ab} \s_{ab} + \hat{\L}^{(1)} + i \hat{\L}^{(1)}_5 \g_5,
\end{equation}
and the vierbein can be generalized to be
\begin{equation}
\hat{e}_\mu = \hat{e}^{(0) a}_\mu \g_a + \hat{e}^{(1) a}_{\mu 5} \g_5 \g_a.
\end{equation}
As a result, additional degrees of freedom appear in the noncommutative case. However, there exists a map, the Seiberg-Witten map \cite{SW},
which relates noncommutative degrees of freedom $\hat{\om}_\mu$, $\hat{e}_\mu$ and $\hat{\L}$
to their commutative counterparts $\om_\mu$, $e_\mu$ and $\L$. For the transformation parameter $\hat{\L}$ and the field $\hat{B}^\mu$,
the map has been derived~\cite{Calmet:2003} up to the first order in $\th^{\mu\nu}$,
\begin{eqnarray}
\hat{\L} &=& \L + \frac{1}{4} \th^{\mu\nu} \{\p_\mu\L, \om_\nu\}, \\
\hat{B^\mu} &=& \th^{\mu\nu} \om_\nu - \frac{1}{4} \th^{\rho\s}\{\om_\rho, \p_\s (\th^{\mu\nu} \om_\nu) + \th^{\mu\nu} R_{\s\nu}\} \label{BMap},
\end{eqnarray}
where $R_{\mu\nu} \equiv \p_\mu \om_\nu - \p_\nu \om_\mu - i[\om_\mu, \om_\nu]$ is the curvature tensor for the spin-connection $\om_\mu$.

The function $\hat{\th}_{\mu\nu} (\hat{X})$ is calculated to the zeroth order in $\th^{\mu\nu}$~\cite{Calmet:2003},
\begin{equation}
\hat{\th}_{\mu\nu} = \th_{\mu\nu} + \th^{\rho\s} \p_\rho \th_{\mu\nu} \om_\s + \mathcal{O} (\th^{\mu\nu}) \label{NCtheta},
\end{equation}
and this is sufficient to compute the curvature tensor $\hat{R}_{\mu\nu}$ up to the first order in $\th^{\mu\nu}$ according to eq.~(\ref{NCRrelation}).
Using eq.~(\ref{Borelation}) and eq.~(\ref{BMap}), we can obtain the map between $\hat{\om}_\mu$ and $\om_\mu$ up to the first order in $\th^{\mu\nu}$,
\begin{equation}
\hat{\om}_\mu = \om_\mu - \frac{1}{4} \th^{\l\s}\{\om_\l, \p_\s \om_\mu + R_{\s\mu}\}
- \frac{1}{4} \th_{\mu\nu} \th^{\l\s} \p_\s \th^{\nu\d} \{\om_\l, \om_\d\} \label{OmMap},
\end{equation}
where the components take the forms,
\begin{eqnarray}
\hat{\om}_\mu^{(0) ab} &=& \om_\mu^{~ab}, \\
\hat{a}^{(1)}_\mu &=& - \frac{1}{16} \th^{\l\s} \om_\l^{~ab} (\p_\s \om_\mu^{~cd} + \frac{1}{2} R_{\s\mu}^{~~cd}) \e_{ac} \e_{bd}
- \frac{1}{16} \th_{\mu\nu} \th^{\l\s} \p_\s \th^{\nu\d} \om_\l^{~ab} \om_\d^{~cd} \e_{ac} \e_{bd}, \\
\hat{b}^{(1)}_{\mu 5} & =& - \frac{1}{32} \th^{\l\s} \om_\l^{~ab}
(\p_\s \om_\mu^{~cd} + \frac{1}{2} R_{\s\mu}^{~~cd}) \ep_{abcd} -
\frac{1}{32} \th_{\mu\nu} \th^{\l\s} \p_\s \th^{\nu\d} \om_\l^{~ab}
\om_\d^{~cd} \ep_{abcd}.
\end{eqnarray}

The Seiberg-Witten map for the vierbein $\hat{e}_\mu$ is
\begin{equation}
\hat{e}_\mu + \d_{\hat{\L}} \hat{e}_\mu = \hat{e}_\mu (e + \d_\L e, \om + \d_\L \om).\label{eSW}
\end{equation}
Thus the solution to the above equation up to the first order in $\th^{\mu\nu}$ can be obtained,
\begin{eqnarray}
\hat{e}_\mu = e_\mu - \frac{1}{2} \th^{\l\s} \{\om_\l, \p_\s e_\mu + \frac{i}{2} [e_\mu, \om_\s]\} \label{eSWMap},
\end{eqnarray}
whose components have the forms,
\begin{eqnarray}
\hat{e}^{(0)a}_\mu &=& e^a_\mu, \\
\hat{e}^{(1)a}_{\mu5} & =& \frac{1}{4} \th^{\l\s} \om_\l^{~eb}
(\p_\s e^c_\mu - \frac{1}{2} \om_\s^{~cd} e_{\mu d})
\ep_{ebc}^{~~~a}.
\end{eqnarray}
Note that eq.~(\ref{eSWMap}) can be verified straightforwardly when it is substituted into eq.~(\ref{eSW}).

Considering the definition eq.~(\ref{SecTensor}) and eq.~(\ref{Borelation}), we get the following expression of $\hat{R}^{\mu\nu}$,
\begin{eqnarray}
\hat{R}^{\mu\nu} &= &\th^{\mu\l} \th^{\nu\s} (\p_\l \hat{\om}_\s - \p_\s \hat{\om}_\l - i [\hat{\om}_\l, \hat{\om}_\s])
+ \frac{1}{2} \th^{\mu\l} \th^{\nu\s} \th^{\d \e} \{\p_\d \hat{\om}_\l, \p_\e \hat{\om}_\s\} \nonumber\\
&& + \frac{1}{2} \th^{\mu\l} \p_\e \th^{\nu\s} \th^{\d\e} \{\p_\d \hat{\om}_\l, \hat{\om}_\s\}
+ \frac{1}{2} \p_\d \th^{\mu\l} \th^{\nu\s}\th^{\d\e} \{\hat{\om}_\l, \p_\e \hat{\om}_\s\}\nonumber\\
&& + \frac{1}{2} \p_\d \th^{\mu\l} \p_\e \th^{\nu\s} \th^{\d \e} \{\hat{\om}_\l, \hat{\om}_\s\}
+ \th^{\s\l} \p_\s \th^{\mu\nu} \hat{\om}_\l + \th^{\mu\nu} - \th^{\mu\nu} (\hat{X}) \label{NCR1}.
\end{eqnarray}
In terms of the Taylor expansion of $\hat{X}(x+\hat{B})$ the last three terms of the above equation are simplified as
\begin{equation}
\th^{\s\l} \p_\s \th^{\mu\nu} \hat{\om}_\l + \th^{\mu\nu} - \th^{\mu\nu} (\hat{X}) = \frac{1}{2} \th^{\s\a}
\th^{\l\b} \p_\s \p_\l \th^{\mu\nu} \hat{\om}_\a \hat{\om}_\b \label{Rtail},
\end{equation}
which contains higher order derivatives of $\th^{\mu\nu}(x)$. When $\th^{\mu\nu} (x)$ is a linear function of $x$ , i.e.
the noncommutativity is of the Lie algebraic structure, eq.~(\ref{Rtail}) vanishes.

In accordance with the Seiberg-Witten map (eq.~(\ref{OmMap})),
the noncommutative curvature tensor $\hat{R}^{\mu\nu}$ in eq.~(\ref{NCR1}) can be expressed
in term of the commutative spin-connection $\om_\mu$,
\begin{eqnarray}
\hat{R}^{\mu\nu} &=&  \th^{\mu\l} \th^{\nu\s} R_{\l\s} + \frac{1}{2} \th^{\mu\l} \th^{\nu\s} \th^{\a\b} \{R_{\l\a}, R_{\s\b}\}
- \frac{1}{4} \th^{\mu\l} \th^{\nu\s} \th^{\s\b} \{\om_\a, (\p_\b+ D_\b) R_{\l\s}\} \nonumber\\
&& + \frac{1}{2} \th^{\mu\l} \p_\a \th^{\nu\s} \th^{\a\b}\{R_{\l\s}, \om_\b\}
+ \frac{1}{2} \p_\a \th^{\mu\l} \th^{\nu\s} \th^{\a\b} \{R_{\l\s}, \om_\b\}
 + \frac{1}{4} \th^{\l\a} \th^{\s\b} \p_\l \p_\s \th^{\mu\nu} \{\om_\a, \om_\b\} ,\nonumber\\
\end{eqnarray}
where $D_\b R_{\l\s} \equiv \p_\b R_{\l\s} - i [\om_\b, R_{\l\s}]$.
With the relation between $\hat{R}^{\mu\nu}$ and $\hat{R}_{\mu\nu}$ (see eq.~(\ref{NCRrelation})) and the expression of
$\hat{\th}_{\mu\nu}$ (see eq.~(\ref{NCtheta})), we obtain $\hat{R}_{\rho\s}$,
\begin{eqnarray}
\hat{R}_{\rho\s} &= &R_{\rho\s} + \frac{1}{2} \th^{\a\b}\{R_{\rho\a}, R_{\s\b}\}
- \frac{1}{4} \th^{\a\b}\{\om_\a, (\p_\b + D_\b) R_{\rho\s}\} \nonumber\\
&& - \frac{1}{2} \th_{\s\nu} \p_\a \th^{\mu\nu} \th^{\a\b} [R_{\rho\mu}, \om_\b]
- \frac{1}{2} \th_{\rho\mu} \p_\a \th^{\mu\nu} \th^{\a\b} [R_{\s\nu}, \om_\b] \nonumber\\
&& + \frac{i}{2} \th^{\a\b} \p_\a \th_{\rho\mu} \p_\b \th^{\mu\nu} R_{\nu\s}
+ \frac{i}{2} \th^{\a\b} \p_\a \th_{\s\nu} \p_\b \th^{\mu\nu} R_{\mu\rho} \nonumber\\
&& + \frac{i}{2} \th^{\a\b} \th^{\mu\nu} \p_\a \th_{\rho\mu} \p_\b R_{\nu\s}
+ \frac{i}{2} \th^{\a\b} \th^{\mu\nu} \p_\a \th_{\s\nu} \p_\b R_{\mu\rho} \nonumber\\
&& + \frac{i}{2} \th^{\a\b} \p_\a \th_{\rho\mu} \p_\b \th_{\s\nu} \th^{\mu\l} \th^{\nu \d} R_{\l\d}
+ \frac{1}{4} \th_{\rho\mu} \th_{\s\nu} \th^{\l\a} \th^{\d\b} \p_\l \p_\d \th^{\mu\nu} \{\om_\a, \om_\b\} \label{NCR2},
\end{eqnarray}
which can also be decomposed by its components as follows:
\begin{equation}
\hat{R}_{\rho\s} = \frac{1}{4} \hat{R}_{\rho\s}^{(0)ab} \s_{ab} +
\frac{1}{2} \hat{R}^{(1)}_{\rho\s} + \frac{i}{2}
\hat{R}^{(1)}_{\rho\s 5} \g_5.
\end{equation}
Therefor the components take the forms,
\begin{eqnarray}
\hat{R}_{\rho\s}^{(0)ab}  &=& R_{\rho\s}^{~~ab} + i \th_{\s\nu} \p_\a \th^{\mu\nu} \th^{\a\b} R_{\rho\mu}^{~~ad} \om_\b^{~cb} \e_{dc}
+ i \th_{\rho\mu} \p_\a \th^{\mu\nu} \th^{\a\b} R_{\s\nu}^{~~ad} \om_\b^{~cb} \e_{dc} \nonumber\\
&& +\frac{i}{2} \th^{\a\b} \p_\a \th_{\rho\mu} \p_\b \th^{\mu\nu} R_{\nu\s}^{~~ab}
+ \frac{i}{2} \th^{\a\b} \p_\a \th_{\s\nu} \p_\b \th^{\mu\nu} R_{\mu\rho}^{~~ab} \nonumber \\
&& + \frac{i}{2} \th^{\a\b} \th^{\mu\nu} \p_\a \th_{\rho\mu} \p_\b R_{\nu\s}^{~~ab}
+ \frac{i}{2} \th^{\a\b} \th^{\mu\nu} \p_\a \th_{\s\nu} \p_\b R_{\mu\rho}^{~~ab} \nonumber\\
&& + \frac{i}{2} \th^{\a\b} \p_\a \th_{\rho\mu} \p_\b \th_{\s\nu} \th^{\mu\l} \th^{\nu\d} R_{\l\d}^{~~ab}, \\
\hat{R}^{(1)}_{\rho\s} &=& \frac{1}{16} \th^{\a\b} R_{\rho\a}^{~~ab} R_{\s\b}^{~~cd} \e_{ac} \e_{bd} 
- \frac{1}{8} \th^{\a\b} \om_{\a}^{~ab} (\p_\b R_{\rho\s}^{~~cd} + \om_\b^{~ec}
R_{\rho\s}^{~~fd} \e_{ef}) \e_{ac} \e_{bd}\nonumber\\
&& + \frac{1}{8} \th_{\rho\mu} \th_{\s\nu} \th^{\l\a} \th^{\d\b} \p_\l \p_\d \th^{\mu\nu} \om_\a^{~ab} \om_\b^{~cd} \e_{ac} \e_{bd},\\
\hat{R}^{(1)}_{\rho\s 5} & =& \frac{1}{32} \th^{\a\b} R_{\rho\a}^{~~ab} R_{\s\b}^{~~cd} \ep_{abcd} 
-\frac{1}{16} \th^{\a\b} \om_\a^{~ab} (\p_\b R_{\rho\s}^{~~cd} + \om_\b^{~ec} R_{\rho\s}^{~~fd} \e_{ef}) \ep_{abcd} \nonumber\\
&& + \frac{1}{16} \th_{\rho\mu} \th_{\s\nu} \th^{\l\a} \th^{\d\b}
\p_\l \p_\d \th^{\mu\nu} \om_\a^{~ab} \om_\b^{~cd} \ep_{abcd},
\end{eqnarray}
where the following identity has been used in the derivation of the above equations,
\begin{equation}
[\s_{ab}, \s_{cd}] = i (\e_{ac} \s_{bd} - \e_{bc} \s_{ad} - \e_{ad} \s_{bc} + \e_{bd} \s_{ac}).
\end{equation}
In terms of the above expression of $\hat{R}_{\rho\s}$ and the vierbein eq.~(\ref{eSWMap}), the action eq.~(\ref{NCaction}) is thus expressed by
the spin-connection $\om_\mu$ and vierbein $e_\mu$ completely,
\begin{eqnarray}
S &=& \int d^4 x \ ({\rm det} \th^{\mu\nu})^{-\frac{1}{2}}
\ep^{\mu\nu\rho\s} \bigg[ (c_0 \e_{ac} \e_{bd} + c_1 \ep_{abcd})
\big(e^a_\mu e^b_\nu \hat{R}_{\rho\s}^{(0)cd}
+ \frac{i}{2} \th^{\a\b} \p_\a (e^a_\mu e^b_\nu) \p_\b R_{\rho\s}^{~~cd}\big)\nonumber\\
& & + c_2 \big(e^a_\mu e^b_\nu e^c_\rho e^d_\s \ep_{abcd} - 2 i
e^a_\mu e^b_\nu e^c_\rho \hat{e}_{\s 5}^{(1)d} \e_{ac}
\e_{bd}\big)\bigg], \label{NCfinalaction}
\end{eqnarray}
where $\hat{R}_{\rho\s}^{(0)cd}$ can be simplified as the following form in terms of the symmetries of indices,
\begin{equation}
\hat{R}_{\rho\s}^{(0)ab} = R_{\rho\s}^{~~ab} + 2 i \th_{\s\nu} \p_\a \th^{\mu\nu} \th^{\a\b} R_{\rho\mu}^{~~ad} \om_\b^{~cb} \e_{dc}
+ i \th^{\a\b} \p_\a \th_{\rho\mu} \p_\b \th^{\mu\nu} R_{\nu\s}^{~~ab} + i \th^{\a\b} \th^{\mu\nu} \p_\a \th_{\rho\mu} \p_\b R_{\nu\s}^{~~ab}.
\end{equation}
Incidentally, the identities below have been used in the calculation of eq.~(\ref{NCfinalaction}),
\begin{eqnarray}
{\rm Tr} (\g_a \g_b) &=& 4 \e_{ab} ,\\
{\rm Tr} (\g_a \g_b \g_5) &=& 0 ,\\
{\rm Tr} (\g_a \g_b \g_c \g_d) &=& 4 (\e_{ab} \e_{cd} - \e_{ac} \e_{bd} + \e_{ad} \e_{bc}) ,\\
{\rm Tr} (\g_a \g_b \g_c \g_d \g_5) &=& -4 i \ep_{abcd}.
\end{eqnarray}


\section{Conclusion}
In this paper, a model of gravity based on the $SL(2,C)$ group is constructed on a noncommutative space with the Poisson structure.
In order to have a covariantly
transformed curvature tensor, a modified function $\hat{\th}_{\mu\nu}$ is introduced. Here,
different from the approach utilized in~\cite{Calmet:2003}
where a kind of covariant coordinates is defined,
we use $\hat{\th}_{\mu\nu}$ to lower the indices of the rank two tensor $\hat{R}^{\mu\nu}$ straightforwardly.
Therefore the gauge invariant action is obtained naturally.
By using the Seiberg-Witten map, we can express the noncommutative physical quantities
in terms of their commutative counterparts and then the action is completely dependent on the commutative quantities $e$ and $\om$.

It is noted~\cite{Calmet:2005, Mukerjee:2006, Banerjee:2007} that the first order correction in actions vanishes.
It is interesting to point out that in our case the first order correction to the Einstein-Hilbert term
in action eq.~(\ref{NCfinalaction}) is a total derivative and thus equals to zero when $\th^{\mu\nu}$ being constant,
which agrees with the result in refs~\cite{Calmet:2005, Mukerjee:2006}.
However, for a general noncommutativity parameter it is evident in our case that the first order correction (see eq.~(\ref{NCfinalaction}))
does not vanish.
The reason lies probably in the different approaches utilized in refs.~\cite{Calmet:2005, Mukerjee:2006, Banerjee:2007} and in the present paper.
In the former approach which is based on the Poincare gauge theory, the vierbein is required to be real. This requirement leads to a
gauge non-invariant action although it preserves the
volume from violation of diffeomorphism in the action.
In the latter approach a $\star$-gauge invariant action is proposed, and it is thus inevitable to introduce a complex vierbein $\hat{e}_\mu$
(see ref.~\cite{Chamseddine:2003} for the canonical noncommutative case).
As a result,
the vanishing first order correction claimed in refs.~\cite{Calmet:2005, Mukerjee:2006, Banerjee:2007} remains in doubt as to whether it happens to
other noncommutative gravity models built in a different way from that of refs.~\cite{Calmet:2005, Mukerjee:2006, Banerjee:2007},
such as to our case.


As a further consideration, we may integrate out $\om_\mu$ in eq.~(\ref{NCfinalaction}) and therefore write the action only in the vierbein.
However, it is quite a challenge to solve the equation
of motion for $\om_\mu$. Moreover, it might be worthwhile to apply our method to other formulations of
noncommutative gravity.

\section*{Acknowledgments}
This work is supported in part by the National Natural Science Foundation of China under grant No.10675061.


\end{document}